\documentstyle[a4,12pt,epsf]{article}

\begin{document}

\title{{\bf Particle emission following Coulomb 
excitation in ultrarelativistic heavy-ion collisions}}
    
\author{ I.A. Pshenichnov$^{1,2,\star}$, I.N. Mishustin$^{2,3}$, 
                             J.P. Bondorf$^{2,\dagger}$,\\
                 A.S. Botvina$^{1,4}$, A.S. Iljinov$^1$\\
{\em $^1$ Institute for Nuclear Research, Russian Academy of Science,}\\
{\em 117312 Moscow, Russia}\\
{\em $^2$ Niels Bohr Institute, DK-2100 Copenhagen, Denmark}\\
{\em $^3$ Kurchatov Institute, Russian Research Center,}\\
{\em 123182 Moscow, Russia}\\
{\em $^4$ Dipartimento di Fisica and INFN, 40126 Bologna, Italy}\\
}
\date{ }
\maketitle

\begin{abstract}
We study nuclear reactions induced by virtual photons associated with
Lorentz-boosted Coulomb fields of ultrarelativistic heavy ions. 
Evaporation, fission and multifragmentation
mechanisms are included in a new RELDIS code, which describes
the deexcitation of residual nuclei 
formed after single and double photon absorption in peripheral heavy-ion
collisions. Partial cross sections for different dissociation channels,
including the multiple neutron emission ones, are calculated and compared
with data when available. Rapidity and transverse momentum distributions of 
nucleons, nuclear fragments and pions, produced electromagnetically, are 
also calculated. These results provide important information for designing 
large-rapidity detectors and zero-degree calorimeters at RHIC and LHC. 
The electromagnetic dissociation of nuclei imposes some constrains on the 
investigation of exotic particle production in $\gamma\gamma$ fusion
reactions. 
\end{abstract}

PACS:  25.75.-q, 25.20.-x, 25.70.De, 24.10.Lx  \\ 
Key words: ultrarelativistic heavy ions, photonuclear reactions, 
Coulomb excitation, intranuclear cascade model, statistical 
multifragmentation\\

$^\star$ E-mail: pshenichnov@al10.inr.troitsk.ru\ \ \ \
$^\dagger$ E-mail: bondorf@nbi.dk

\newpage

\section{Introduction}
 
There exist many mechanisms of particle 
production, nucleon and fragment emission in nuclear collisions at
ultrarelativistic energies.  
They  can be divided into the following ample categories.  
The first category contains collisions with direct 
nuclear overlap. In addition to nucleon-nucleon or parton-parton collisions, 
which are the most obvious sources of secondary hadrons, 
collective mechanisms were 
considered for baryon-antibaryon pair~\cite{Mishustin1} or 
dilepton~\cite{Mishustin2} production.  
The second category deals with
"distant" electromagnetic collisions without nuclear overlap. In such a 
situation the interaction can be treated in terms of  
virtual photons emitted by one nucleus and absorbed by another collision 
partner. A subsequent photonuclear ($\gamma A$) reaction leads to a so-called
Electromagnetic Dissociation (ED) of nuclei~\cite{Winther,BertBaur}. 
The virtual photon flux and consequently the rate of such $\gamma A$ events 
is proportional to $Z^2$, where Z is the charge of the first nucleus.  
The ED process is widely used for studying the properties of Giant Resonances 
(GR) in nuclei, including multiphonon excitations~\cite{BertBaur}.

In Ref.~\cite{Iljinov}  the description of nuclear photoabsorption was extended
to the photon energies much above the GR region, where the exitation of 
individual nucleons and multiple pion production become important.   
The model of electromagnetic dissociation processes taking into account these 
high-energy photon absorption channels was constructed in 
Ref.~\cite{Pshenichnov}.
According to this model, the fast hadrons produced after the photon absorption 
initiate a cascade of subsequent collisions with the intranuclear nucleons 
leading to the heating of a residual nucleus.   
At a later stage the nucleus undergoes de-excitation by means of the 
evaporation of nucleons and lightest fragments, binary fission or 
multifragmentation. The latter process takes place when the excitation energy 
is comparable with the total binding energy of a residual nucleus~\cite{JPB}. 
 
Another interesting mechanism in the second category is the 
photon-photon collision or $\gamma\gamma$ fusion leading to the particle 
production at mid-rapidity~\cite{MGreiner,BaronBaur,Vidovic}. 
The rate of these processes is proportional to $Z^4$. 
The effective $\gamma\gamma$ luminosity at the LHC collider is expected 
to be comparable with that at a possible future Photon Linear 
Collider~\cite{HenckenTrautmann}.

It is worthwhile to mention that 
the processes of the second category have attracted a special attention in 
the last few years. 
This interest is generated by the 
construction of the new heavy-ion colliders: RHIC at Brookhaven
(${\rm Au}+{\rm Au}$  with $\sqrt{s}=200$ GeV) and LHC at CERN  
(${\rm Pb}+{\rm Pb}$  with $\sqrt{s}=5500$ GeV). It is expected that
the ED processes will play an important role for these energies and
especially for the ions with high $Z$, due to the above-mentioned 
$Z^2$-dependence.  The predicted cross section for the electromagnetic 
dissociation at LHC, $\sigma_{ED}$, is so large ($\approx 200$ b) 
that together with the electron capture process ($\approx 100$ b), 
it will reduce significantly  the life time of relativistic 
heavy-ion beams~\cite{ALICE} as compared 
with the proton ones. The ED reaction rate at LHC is expected to be 
30 times higher than that for the nuclear interactions 
($\sigma_{nuc}\approx 7$ b). But of course the ED reactions are much
less violent then the nuclear ones.    

Hadron production in photon-photon and photon-pomeron  collisions, 
as well as in double-pomeron exchange were studied in Ref.~\cite{Engel}. 
All these
reactions are characterised by a large rapidity gap separating the remnants of 
the colliding hadrons or heavy ions from the particles produced in the central
rapidity region. High-mass diffraction reactions were simulated by means of the 
Monte-Carlo event generator Phojet~\cite{Engel2}. On the other hand, there are
no "ready-to-use" event generators for simulating relatively 
``soft'' $\gamma A$ processes, where nuclear degrees of 
freedom play an important role. In particular, such simulations are 
needed for estimating the probabilities of different disintegration 
channels of the colliding nuclei.
The properties of the ED processes have to be well understood in order to 
distinguish them from very peripheral nuclear collisions.

In the present paper we consider the whole set of mechanisms of Coulomb 
excitation and subsequent dissociation of relativistic heavy-ions (RHI). 
In Sec.2 the model is presented and compared with existing experimental
data on few neutron removal from Au and Pb nuclei by virtual and 
real photons. 
In Sec.3 we discuss the ED processes which are
expected to be important at the future heavy-ion colliders, RHIC and LHC.  
We present the model predictions for
multiplicity and transverse momentum distributions of neutrons. These results
are important for designing zero-degree calorimeters for 
beam luminosity monitoring. 
Rapidity and transverse momentum distributions of $\pi^+$ and $\pi^-$ mesons,
fission products and intermediate mass fragments are also discussed in Sec.3. 
Sec.4 is reserved for discussion and conclusions.

\section{Model of electromagnetic dissociation of heavy nuclei}

\subsection{Single- and double-photon absorption}

Let us consider the Coulomb excitation of an ultrarelativistic projectile.
According to the Weizs\"{a}cker-Williams (WW) method
the spectrum of virtual photons from a stationary target of charge $Z_t$
as seen by a projectile moving with velocity $\beta =v/c$ 
at impact parameter $b$ is expressed as~\cite{JDJackson}:
\begin{equation}
N(E_\gamma ,b)=\frac{\alpha Z^{2}_t}{\pi ^2}
\frac{{\sf x}^2}{\beta ^2 E_\gamma b^2} 
\Bigl(K^{2}_{1}({\sf x})+\frac{1}{\gamma ^2}K^{2}_{0}({\sf x})\Bigl),
\label{eq:1}
\end{equation}
where $\alpha$ is the fine structure constant,
$K_0$ and $K_1$ are the modified Bessel functions of zero and first
order, ${\sf x}=E_\gamma b/(\gamma \beta \hbar c)$.
It is worth noting that the distribution (\ref{eq:1}) 
is written in the projectile rest frame. This means that for
colliding beams one should use the Lorentz factor of one collision
partner in the rest frame of another partner, $\gamma =2\gamma_{beam}^2-1$,
where $\gamma_{beam}$ is the Lorentz factor of each beam.

The projectile can be exited by absorbing one or more 
virtual photons.  Multiple excitations of the giant dipole resonance (GDR) up
to the fourth order were considered in Refs.~\cite{BertBaur,Llope}.
Since the $nth$-order cross section behaves approximately as 
$(\alpha Z_{t}^{2})^{n}$, one
should expect the maximum effect for
the excitation of $^{238}{\rm U}$ projectile in the field of $^{238}{\rm U}$
target at ultrarelativistic energies.
As was found in Ref.~\cite{BertBaur}, even in 
this case the cross sections of the third and fourth order processes are 
quite small in comparison with that for the second order
excitation: $\sigma^{(3)}/\sigma^{(2)}\approx 0.11$ and 
$\sigma^{(4)}/\sigma^{(2)}\approx 0.01$.
Thus we restrict our consideration by the first and second order processes. 

The mean number of photons absorbed by the projectile of mass $A_p$ 
in a collision at impact parameter $b$ is defined by
\begin{equation}
m(b)=
\int\limits_{E_{min}}^{\infty}N(E_\gamma ,b)\sigma_{A_p}(E_\gamma)dE_\gamma ,
\label{eq:2}
\end{equation}
where $\sigma_{A_p}(E_\gamma)$ is the appropriate photoabsorption cross
section, either measured for this projectile nucleus with real photons
or calculated within a model.
For the cases of interest, i.e. 
for Au and Pb nuclei, we use the data set published in 
Refs.~\cite{Veyssiere,Lepretre,Harvey,Berman}, including a most recent 
one~\cite{Mirazita} 
where corresponding photoabsorption cross sections were measured. 
In the GDR region one can use a set of Lorentz-line 
approximations  given in the review~\cite{Berman-Fultz}.  
According to prescriptions
of Ref.~\cite{Berman}, the photoabsorption cross sections obtained
in Refs.~\cite{Veyssiere} and ~\cite{Harvey} for Au and Pb nuclei are rescaled 
by factors 0.93 and 1.22,
respectively. 

Following Llope and Braun-Munzinger~\cite{Llope} we assume that at an impact
parameter $b$ the probability of the multi-photon absorption is given
by the Poisson distribution with the mean multiplicity 
$m(b)$ defined by Eq.~(\ref{eq:2}). 
Then the probability
for absorbing exactly one photon of any energy during a collision
at impact parameter $b$ is equal to:
\begin{equation}
P^{(1)}(b)=m(b)e^{-m(b)},  
\label{eq:3}
\end{equation}
and the probability for absorbing exactly two photons is:
\begin{equation}
P^{(2)}(b)=\frac{m^{2}(b)}{2!}e^{-m(b)}.  
\label{eq:4}
\end{equation}

The normalized probability density that in a collision at impact parameter 
$b$ the absorbed photon has an energy $E_1$ is given by:
\begin{equation}
q^{(1)}(E_1,b)=\frac{N(E_1,b)\sigma_{A_p}(E_1)}{m(b)}.
\label{eq:5}
\end{equation}
Analogously, the probability density that in a  
second order process two photons have energies $E_1$ and $E_2$ is:
\begin{equation}
q^{(2)}(E_1,E_2,b)=
\frac{N(E_1,b)\sigma_{A_p}(E_1)N(E_2,b)\sigma_{A_p}(E_2)}{m^{2}(b)}.
\label{eq:6}
\end{equation}
 
To consider particular ED channels i.e. neutron emission, fission, 
multifragmentation or pion production, which are studied in 
the present paper, we define the 
differential cross sections for the first and 
second order processes as:
\begin{equation}
\frac{d\sigma^{(1)}_{i}}{dE_1}=2\pi
\int\limits_{b_{min}}^{\infty}bdbP^{(1)}(b)q^{(1)}(E_1,b)f^{(1)}_{i}(E_1)
\label{eq:7}
\end{equation}
and 
\begin{equation}
\frac{d^{2}\sigma^{(2)}_{i}}{dE_1dE_2}=2\pi
\int\limits_{b_{min}}^{\infty}bdbP^{(2)}(b)q^{(2)}(E_1,E_2,b)
f^{(2)}_{i}(E_1,E_2)
\label{eq:8}
\end{equation}
Here $f^{(1)}_{i}(E_1)$ and $f^{(2)}_{i}(E_1,E_2)$
are the branching ratios for the considered channel~$i$.

Finally, the integral cross sections are calculated as:
\begin{equation}
\sigma^{(1)}_{i}=
\int\limits_{E_{min}}^{\infty} dE_1 N^{(1)}(E_1)
\sigma_{A_p}(E_1)f^{(1)}_{i}(E_1),
\label{eq:9}
\end{equation}
\begin{equation}
\sigma^{(2)}_{i}=
\int\limits_{E_{min}}^{\infty}\int\limits_{E_{min}}^{\infty}dE_1dE_2 
N^{(2)}(E_1,E_2)\sigma_{A_p}(E_1)\sigma_{A_p}(E_2)
f^{(2)}_{i}(E_1,E_2).
\label{eq:10}
\end{equation}
Here we introduce the single and double photon spectral functions
which appear after 
the integration over impact parameters;
\begin{equation}
N^{(1)}(E_1)=2\pi\int\limits_{b_{min}}^{\infty} bdb e^{-m(b)} N(E_1,b),
\label{eq:11}
\end{equation}
\begin{equation}
N^{(2)}(E_1,E_2)=\pi\int\limits_{b_{min}}^{\infty} bdb 
e^{-m(b)} N(E_1,b) N(E_2,b).
\label{eq:12}  
\end{equation}

In the above expressions $b_{min}$ is the minimal value
of the impact parameter which corresponds to the onset of nuclear interaction.
The integration over the photon energy starts from 
$E_{min}\approx 7$ MeV which corresponds to the threshold of nuclear 
dissociation in photonuclear reactions. Due to the long-range nature of the 
electromagnetic forces, numerical calculations of the integrals~(\ref{eq:11}) 
and~(\ref{eq:12}) should be undertaken with care. 
A reasonable accuracy can be obtained 
if one splits the whole region of integration into two intervals:
$[b_{min},b_{cut}]$ and $[b_{cut},\infty)$. The value of $b_{cut}$ is
determined by the condition $e^{-m(b_{cut})}\approx 1$ which allows one to omit
the exponential factor in the second interval.  
After that the first integral in Eq.~(\ref{eq:11})
is calculated numerically, while the second one can be evaluated
analytically, leading to a well-known expression~\cite{JDJackson}.
The same splitting can be made in the integral~(\ref{eq:12}). Using the  
asymptotic behaviour at ${\sf x}\gg 1$:
\begin{equation} 
K_1({\sf x})\approx\sqrt{\pi/2{\sf x}}e^{-{\sf x}},    
\end{equation}    
one obtains the exponential integral which can be found in  
mathematical tables. With the described method one can avoid problems
with too many nodes
in the numerical integration over the wide range of impact parameters.  

The values of $f^{(1)}_{i}$ and $f^{(2)}_{i}$ for branching ratios of 
partial channels of photonuclear dissociation induced by real and virtual 
photons were calculated within the model of Ref.~\cite{Pshenichnov}  which 
takes into account the 
fast cascade stage of the reaction  as well as  evaporation, 
fission and multifragmentation processes for a residual nucleus.

\subsection{Excitation and decay of residual nuclei}

Relative probabilities of the de-excitation processes are determined  
by the excitation energy $E^\star$ of a residual nucleus formed 
after the completion of the
fast cascade stage of the photonuclear reaction. In the following we 
consider step-by-step all these mechanisms with a special attention to 
the amount of energy which is transformed on average into the internal 
excitation of the nucleus.   

Depending on the virtual photon energy $E_\gamma$ different processes
may contribute to the energy deposition. 
When a nucleus absorbs one or two virtual photons in the GR region, 
$6\leq E_\gamma \leq 30$ MeV,  
their energies are completely transformed into the 
excitation energy $E^\star$. For preactinide nuclei like $\rm Au$ and 
$\rm Pb$, whose fission thresholds are about 30 MeV, the 
de-excitation proceeds mainly through the evaporation of neutrons, since 
their separation energies are only around 7 MeV. Due to a high Coulomb
barrier in heavy nuclei, the proton emission is suppressed in the GR region. 
A detailed experimental investigation of 
$(\gamma ,n)$, $(\gamma ,2n)$, $(\gamma ,3n)$, and $(\gamma ,4n)$ reactions
on $\rm^{197}Au$ and $\rm^{208}Pb$ nuclei in the GR region was performed in 
Ref.~\cite{Veyssiere}. 
The presence of the  reaction $(\gamma ,4n)$ means that a nucleus has 
received $E^\star$ up to 30 MeV.  For simulating the particle emission
in the evaporation process 
we used the standard Weisskopf evaporation scheme~\cite{JPB} with the assumption
$E^\star =E_\gamma$.
  
Starting from $E_\gamma =$ 30 MeV, where the quasideuteron mechanism becomes
important and up to the single pion production threshold at $E_\gamma =140$ 
MeV, only a part of the photon energy is converted on average  into the 
excitation energy $E^\star$ of the compound nucleus. The rest of the photon 
energy is taken away by the fast nucleons originating from the absorbing pair. 
The energy deposition is simulated by the IntraNuclear Cascade (INC) 
model (see Refs.~\cite{Iljinov,Pshenichnov} for details).  
The two-nucleon absorption cross section of a
photon on a heavy nucleus, $\sigma_A$, is taken from the  
quasideuteron model of Ref.~\cite{Lepretre}:
\begin{equation}
\sigma^{QD}_A = k Z (1-Z/A) \sigma_d^{exch}.
\label{eq:13}
\end{equation}
\noindent
Here $\sigma_d^{exch}$ is the meson exchange part of the cross section for 
the deuteron photodisintegration, $\gamma d \rightarrow np$.
This cross section was calculated in Ref.~\cite{Laget}. In Eq.~(\ref{eq:13})
$A$ and $Z$ are the mass and charge numbers of the relevant nucleus 
and $k\approx$11~\cite{Lepretre} is an empirical constant. 
The angular distribution of nucleons in the reaction
$\gamma d \rightarrow np$ was approximated in accordance with empirical fits
from  Ref.~\cite{Barashenkov}. Although the cross
section $\sigma_d$ decreases strongly with the photon energy, the two-nucleon
absorption mechanism competes noticeably  with the single-nucleon 
photoabsorption channel up to $E_\gamma \sim$0.5 GeV. 

As found in Ref.~\cite{Lepretre}, up to 12 neutrons can be emitted by  
photo-excited $\rm Pb$ nucleus. 
Assuming that the emission comes from the compound nucleus one can conclude 
that $\langle E^\star\rangle \leq 70-80$ MeV.
The calculations of Ref.~\cite{Guaraldo} for
photoabsorption on $\rm^{197}Au$ and $\rm^{208}Pb$ nuclei, performed
in the framework of another version of INC model~\cite{Barashenkov}, 
demonstrate that $\langle E^\star\rangle $ does not exceed 80 MeV
for $30\leq E_\gamma\leq 140$ MeV. 
In this $E_\gamma$ region the fissility $P_f$ of $\rm^{197}Au$ 
and $\rm^{208}Pb$ nuclei becomes noticeable ($P_f\sim 0.01$ and $0.1$, respectively),
so one should expect the evaporation-fission competition. 

Above the single pion production threshold at $E_\gamma =140$ MeV  
the $\Delta$-isobar excitation of individual nucleons dominates. 
An interesting effect was noticed in Ref.~\cite{Guaraldo}. 
Near the threshold a slow pion has a small interaction cross 
section with nucleons and therefore,
has a high probability to carry away a large part ($\approx m_\pi$) of the photon
energy. Only at $E_\gamma\approx 200$ MeV the average value 
$\langle E^\star\rangle $ 
starts to increase again. During the fast cascade stage the initial nucleus looses
on average 1-2 nucleons and acquires the excitation energy 
$\langle E^\star\rangle \sim 100$ MeV, 
which is sufficient for evaporation of many neutrons or fission. 
The evaporation may take
place before of after the fission. In some cases fission fragments evaporate
neutrons as well.

Above the two-pion production threshold at 
$E_\gamma\sim 400$ MeV the photon-nucleon ($\gamma N$) interaction becomes 
more complicated because of many possible final states. 
Below we use a phenomenological model for exclusive description
of the $\gamma N$-interaction developed in Ref.~\cite{Iljinov}. The model 
includes both the resonance
contribution from the two-body channels, $\gamma N \rightarrow \pi B^\star$
and $\gamma N \rightarrow M^\star N$ ($B^\star$ and $M^\star$ being baryon
and meson resonances), and the non-resonant statistical contribution from the
multi-body channels $\gamma N \rightarrow i \pi N$ ($2 \leq i \leq 8$). 
Subsequently, the fast hadrons produced in a primary $\gamma N$-interaction
initiate a cascade of successive hadron-nucleon collisions inside 
the nucleus. Finally, when the photon
energy reaches the value of several GeV, the multiple pion production
becomes the main process. A large number ($\sim 80$) of many-body
subchannels is included in the calculation.

\begin{figure}[hp]  
\begin{centering}
\epsfxsize=0.75\textwidth
\epsffile{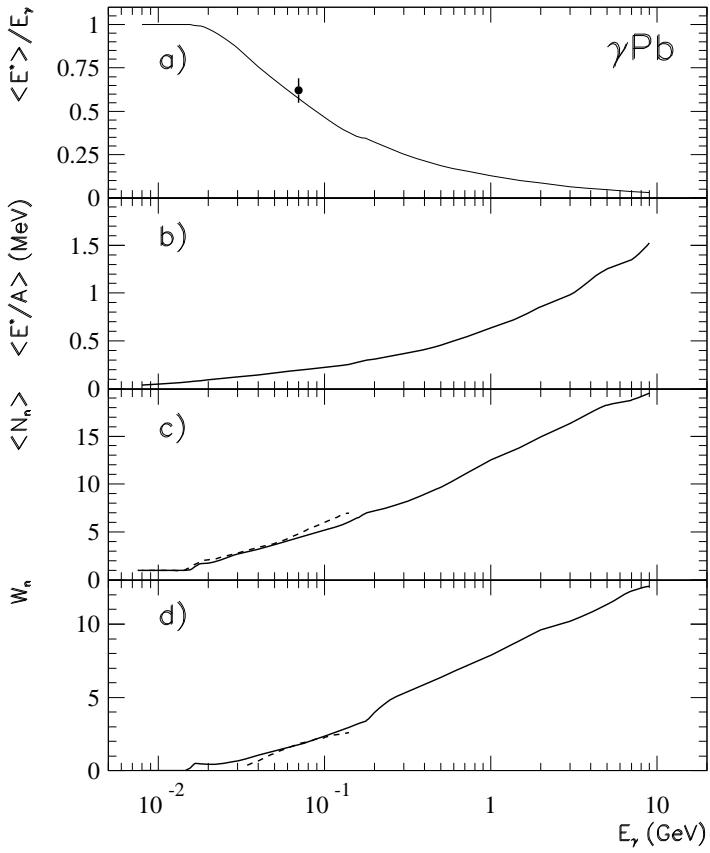}
\caption{a) Ratio of average excitation energy of residual nucleus $E^\star$
to input photon energy $E_\gamma$, b) absolute values of $E^\star$ per
nucleon of residual nuclei, c) average photoneutron multiplicities 
$\langle N_n\rangle $ and
d) width of photoneutron multiplicity $W_n=
\protect\sqrt{\langle N_n^2\rangle -\langle N_n\rangle ^2}$   
as functions of $E_\gamma$ in photoabsorption on lead nucleus. Solid lines 
show the RELDIS
calculations, dashed lines and point represent 
the experimental data~\protect\cite{Lepretre3}.
}
\label{fig:0}
\end{centering}
\end{figure}

Fig.~\ref{fig:0} shows the fraction of $E_\gamma$ that on average is 
transformed into $E^\star$ as well as average values of $E^\star$ per nucleon 
of residual nucleus. As it was deduced from the experimental 
data~\cite{Lepretre3}, $E^\star = 43.4\pm5$ MeV  for 
photoabsorption on lead at $E_\gamma=70$ MeV. This value agrees well with the 
prediction of our model (see Fig.~\ref{fig:0}a). In Fig.~\ref{fig:0}a one can
see the above-mentioned decrease of $\langle E^\star\rangle /E_\gamma$ 
above $E_\gamma=140$ MeV. As seen from Fig.~\ref{fig:0}, when the virtual 
photon energy grows from the GR region to several GeV, the nature of the 
Coulomb excitation changes from the collective nuclear degrees of 
freedom to the excitation of a single nucleon inside the nucleus. In the
latter case up to 95\% 
of the photon energy is released in the form of fast particles
leaving the nucleus. Nevertheless, the remaining  energy deposited 
in the residual nucleus is sufficient
for evaporating many neutrons.  The moments of the 
neutron multiplicity distribution are shown in Fig.~\ref{fig:0},
parts c and d. Calculated moments are in a satisfactory
agreement with the experimental values for the real photon absorption on Pb
nucleus~\cite{Lepretre3}. Below we use the same model for the case of
the virtual photon absorption in electromagnetic
interactions of ultrarelativistic heavy ions.    

It should be stressed however, that 
Fig.~\ref{fig:0} shows only the average values, while a part of the 
interactions initiated by "hard" photons will lead to a stronger heating of 
the nucleus by a multi-pion system. This phenomenon was studied in detail 
in Ref.~\cite{Pshenichnov} were it was shown that the yield of events 
with high excitation energy, resulting 
in the multifragment break-up of residual nuclei, might be
quite sizable, up to $10\div15\%$ of the total ED cross section. 
These events can be easily observed with properly designed
forward-angle detectors at RHIC and LHC colliders. 

To include all the processes
described above we have developed a specialized computer code RELDIS
aimed at the Monte Carlo simulation of the Relativistic ELectromagnetic 
DISsociation of nuclei. The simulation begins 
with generating the single or double-photon absorption process. Then the 
intranuclear cascade model is used to calculate the fast particle emission
and the characteristics of residual nuclei. Finally the de-excitation of 
thermalized residual nuclei is simulated according to the statistical
model \cite{JPB}.

\subsection{Analysis of nucleon emission data}

Let us consider now how the calculations based on 
Weizs\"{a}cker-Williams (WW) method describe  
the measured~\cite{Aumann1} cross sections of $1n$, $2n$ and $3n$ removal 
from $^{197}{\rm Au}$ nuclei excited electromagnetically.
The branching ratios, $f^{(1)}_{i}$ and $f^{(2)}_{i}$, for partial channels  
of a photonuclear reaction are important ingredients of such calculations. 
The information concerning these ratios can be obtained from 
experiments with real 
photons~\cite{Veyssiere,Berman,Lepretre3} or from models of photonuclear 
reactions analogous to the one used in the present paper. We review in brief  
the advantages and disadvantages of both these approaches. 

The calculations~\cite{Aumann1} within the framework of 
WW method, which include both
isoscalar and isovector quadrupole excitations as well as multiphonon 
states,  describe the one-neutron emission data fairly well. Using the 
harmonic-oscillator model one can derive a Poisson distribution for the
excitation probabilities of multiphonon states.  
On the contrary, even the theory with these additional modifications 
is not able to describe the $2n$ removal cross section.
The measured cross section is twice as large as the 
calculated value, which is determined mainly by the probability of the
double-phonon excitation.  Further modifications of the WW method, using
a "soft-spheres" model and including nuclear-plus-Coulomb
processes~\cite{Aumann2}, do not remove the discrepancy. 
The description of the reaction mechanism within the coupled-channel 
approach~\cite{Bertulani2}  with proper account for the nuclear interaction 
is not able to improve the situation. Only the inclusion  of anharmonicities 
in the internal Hamiltonian and nonlinear terms in the external
field, proposed in Ref.~\cite{Volpe} and later supported  by 
other authors~\cite{Bortignon}, help to solve the
problem. As shown in Ref.~\cite{Volpe,Bortignon}, a small deviation from the 
harmonic model may lead to a doubling of the two-phonon excitation 
probability and to a good description of the data.

The calculations of Refs.~\cite{Volpe,Bortignon} are based on the 
one- and two-neutron emission cross sections measured in experiments with real 
photons. Thus only a limited number of channels of electromagnetic 
dissociation was considered.
The differences in the total GDR cross section measured by different groups 
are found to be on the level of 20\%~\cite{Berman}. 
There exists also an opinion~\cite{Wolynec} that for the $2n$ photoemission 
measurements the quality of the data is even much worse. Different 
interpretations of the photonuclear data lead to difficulties in 
determining the contribution of the double-photon absorption.  
Indeed, the double GDR state may decay via $2n$ and $3n$ channels and 
therefore its admixture depends on the corresponding partial cross sections.
 
In the present paper we analyse the ED data~\cite{Aumann1}  
within the model of photonuclear reactions introduced above. 
We {\em do not use} experimental data as an input for calculations, except
of the information concerning the non-statistical contribution in the $1n$ 
emission channel.
It should be reminded, that the statistical decay of the compound nucleus 
is not the only mechanism responsible for the decay of the collective GDR 
state. This state can also decay directly into a free nucleon and a hole 
in the residual nucleus, due to the two-body residual interaction in the 
final state (see Refs.~\cite{Veyssiere,Van} for details). Such a direct 
neutron emission leaves the residual nucleus in a low-lying energy state 
where only a subsequent emission of photons is possible.   
This makes the emission of a second neutron impossible, leading to a 
suppression of the $(\gamma ,2n)$ channel in comparison with the pure 
statistical decay.
We include this additional non-statistical 
contribution in the $(\gamma ,n)$ channel on the basis of the 
experimental information concerning this decay mode.  
The total fraction of the non-statistical contribution was evaluated to 
about $35\% $ and $20\% $ for Au and Pb nuclei, 
respectively~\cite{Veyssiere}. We use these
values in our simulations. The emission angles  of non-statistical fast 
photoneutrons were generated according to the angular distributions 
measured in Ref.~\cite{Tagliabue}.      
 
In our model of photonuclear reactions, implemented in the RELDIS code, 
we consider all possible decay modes, including binary fission and 
multifragmentation. It gives the yields and spectra of different 
particles in the ED reactions induced by real and virtual photons.
Before discussing the predictions in the remaining part of the paper, 
let us consider first how the model describes the existing data on a few 
neutron removal from Au nuclei~\cite{Aumann1}. 

The results for the few-nucleon emission are summarized in Table~\ref{T1}.
One can see that the model describes the $3n$ 
emission quite well but the agreement with the measured one- and 
two-neutron emission cross sections is not as good. At the same time the 
total neutron 
$(n-3n)$ emission cross section agrees with the experimental value.  
Due to the statistical
nature of our model one can expect that it is not so accurate for
the particle emission from collective states like GDR, 
but works better for the multiple emission from the compound nucleus.
This tendency is clearly seen in Table~\ref{T1}. The higher is the energy
of colliding ions the greater is the number of emitted particles. 
Therefore, the quality of description of the low energy data~\cite{Aumann1} 
may be considered as a low limit for the accuracy of the model which is 
expected to work better with increasing  energy. It should be noticed
that a simple Poisson distributions for multiple excitations  was adopted in 
our calculations.  And even within this assumption the agreement with 
data~\cite{Aumann1} is acceptable, in contrast with observations 
of Refs.~\cite{Aumann1,Aumann2}. This discrepancy is due to slightly 
different values of the partial photonuclear cross sections ($2n$ and $3n$) 
used in our calculations as compared with the papers~\cite{Aumann1,Aumann2}.
            
The importance of the double-photon absorption 
mechanism is illustrated in Tab.~\ref{T1} too. Although the total 
cross section of double-photon absorption, $\sigma^{(2)}=226$ mb, 
is small compared to the single photon one, $\sigma^{(1)}=3546$ mb, 
this contribution is very significant for the rare channels like
$(\gamma ,4n)$, $(\gamma ,5n)$, $(\gamma ,6n)$ and, especially, for
$(\gamma ,p3n)$, $(\gamma ,p4n)$ and $(\gamma ,p5n)$ channels. These listed 
channels are extremely sensitive to the presence of the double photon
absorption mechanism.

\section{Predictions for ED at RHIC and LHC}

\subsection{Total and partial ED cross sections}

Let us consider now the model predictions for the electromagnetic
dissociation of ultrarelativistic Pb and Au nuclei on fixed targets at SPS
and in colliding beams at RHIC and LHC. 
The total ED cross section, $\sigma_{ED}=\sigma^{(1)}+\sigma^{(2)}$,
is obtained from Eqs.~(\ref{eq:9}) and (\ref{eq:10}) by taking 
$f^{(1)}_{i}=1$ and $f^{(2)}_{i}=1.$
The $\sigma_{ED}$ values for most interesting reactions are presented 
in Tab.~\ref{T2}. 

\begin{figure}[hp]  
\begin{centering}
\epsfxsize=0.85\textwidth
\epsffile{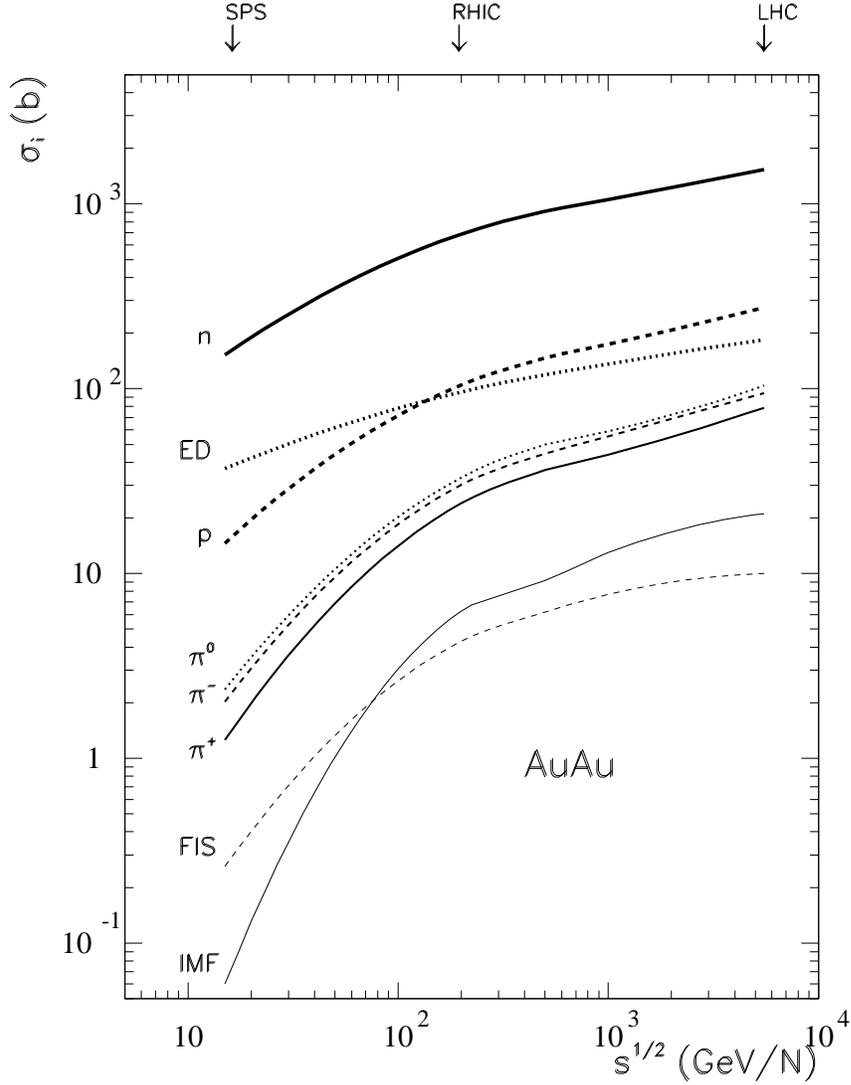}
\caption{RELDIS predictions for inclusive cross sections of 
emission of nucleons, nuclear fragments and pions in the electromagnetic 
dissociation of {\rm one} of the colliding Au nuclei.
Cross sections for 
intermediate mass ($3\leq Z\leq 30$) and fission  ($30<Z\leq 50$)  
fragments are labelled as ``IMF'' and ``FIS'', respectively. Thick dotted
line shows the total electromagnetic dissociation cross section.
$\protect\sqrt{s}$ values for SPS, RHIC and LHC are shown by arrows.}
\label{cs}
\end{centering}
\end{figure}

The inclusive (multiplicity weighted) cross sections for emission of 
nucleons, nuclear fragments and pions 
in the electromagnetic dissociation 
of {\it one} of the colliding ${\rm Au}$ nuclei are shown in 
Fig.~\ref{cs} as functions of the beam energy. 
In addition to Fig.~\ref{cs}, where the calculations are 
presented only for ${\rm Au+Au}$ reaction, in Tab.~\ref{T3} the inclusive cross 
sections are given for the three reactions listed in Tab.~\ref{T2}. 
One can see that
the partial cross sections for neutron emission are especially large.
This is due to the high average neutron multiplicities, 
4.1, 7.2 and 8.8, 
at SPS, RHIC and LHC energies, respectively. 

The relative contributions to the inclusive cross sections
from the double photon absorption, 
$\sigma^{(2)}_{i}/(\sigma^{(1)}_{i}+\sigma^{(2)}_{i})$,
are listed for the same reactions in Tab.~\ref{T4}.
By inspecting Tab.~\ref{T4} one can conclude that about 10\% , 5\% and 2\% 
of particles, on average, are produced in the second order processes 
at SPS, RHIC and LHC energies, respectively. 
These values should be compared with the corresponding contributions
to the total ED cross sections (Tab.~\ref{T2}), 
$\sigma^{(2)}/\sigma_{ED}$, which are, 3.5\%, 2.3\% and 1.4\%, 
respectively.  
Thus, the naive expectation, that
in the double photon absorption the multiplicities of particles should be
exactly twice as large as in the single photon absorption, is not valid. 
Indeed, according to this expectation the partial fractions should be 
twice as large as the fractions of $\sigma^{(2)}$ in  the total cross 
section, i.e. 7\% , 4.6\% and 2.8\% . 
This rule is violated for the intermediate mass and fission 
fragments. Nevertheless the general tendency is obvious: high order 
corrections become less and less important with increasing 
energy. This is true not only for the total ED cross section, as
it was recently noticed in Ref.~\cite{Norbury}, but also for the 
partial cross sections for nucleon and fragment emission and pion 
production.

\subsection{Neutron multiplicity distributions}

As stressed above, neutrons are the most abundant particles produced
in the ED process. In addition, there are many mechanisms of neutron emission
that makes the calculation of their characteristics especially difficult.
During the fast stage of the photonuclear reaction some intranuclear
nucleons may escape from the nucleus via a direct knock-out either by the
incoming photon or by a secondary particle in the course of the intranuclear 
cascade. In addition, as it was mentioned before, many neutrons may be 
evaporated from an excited residual nucleus. 

\begin{figure}[hp]  
\begin{centering}
\epsfxsize=0.95\textwidth
\epsffile{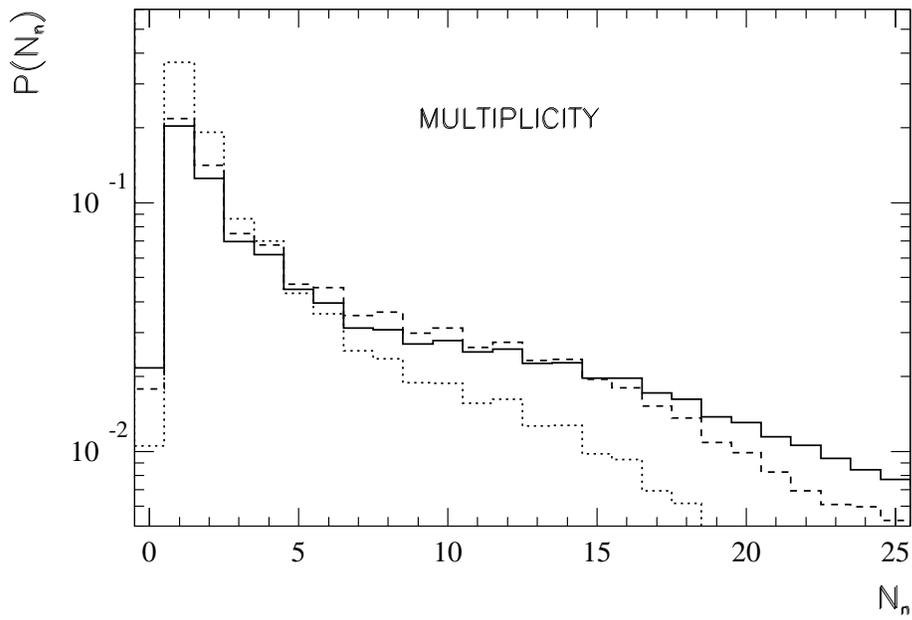}
\caption{ RELDIS predictions for multiplicity distributions of 
neutrons in the electromagnetic dissociation 
of Pb nuclei at LHC and SPS energies (solid and dotted lines,
respectively) and Au nuclei at RHIC energies (dashed lines).}
\label{fig:nmult}
\end{centering}
\end{figure}

The neutron multiplicity distributions predicted by the RELDIS code 
are presented in Fig.~\ref{fig:nmult} for the three reactions discussed above.
One can see that these distributions have a non-trivial structure.
They are strongly peaked at the ${1n}$ 
emission channel associated with the GR decay. On the other hand, there is 
a long tail of multiple neutron emission associated  with the knock-out and
evaporation processes. 
One can see, for example, that the probability to emit more than 
20 neutrons is quite noticeable ($\approx 5\%$ at RHIC).
The mean neutron multiplicities for the three reactions considered are 
4.2, 7.2 and 8.8 at SPS, RHIC and LHC energies, respectively.
These results might be important for designing zero-degree calorimeters at 
RHIC and LHC.

\subsection{Spectra of nucleons, pions and nuclear fragments}

Now let us discuss
the differential distributions in rapidity $y_{lab}$ and transverse 
momentum $P_t$ for  nucleons, nuclear fragments and pions.   

\begin{figure}[hp]  
\begin{centering}
\epsfxsize=0.95\textwidth
\epsffile{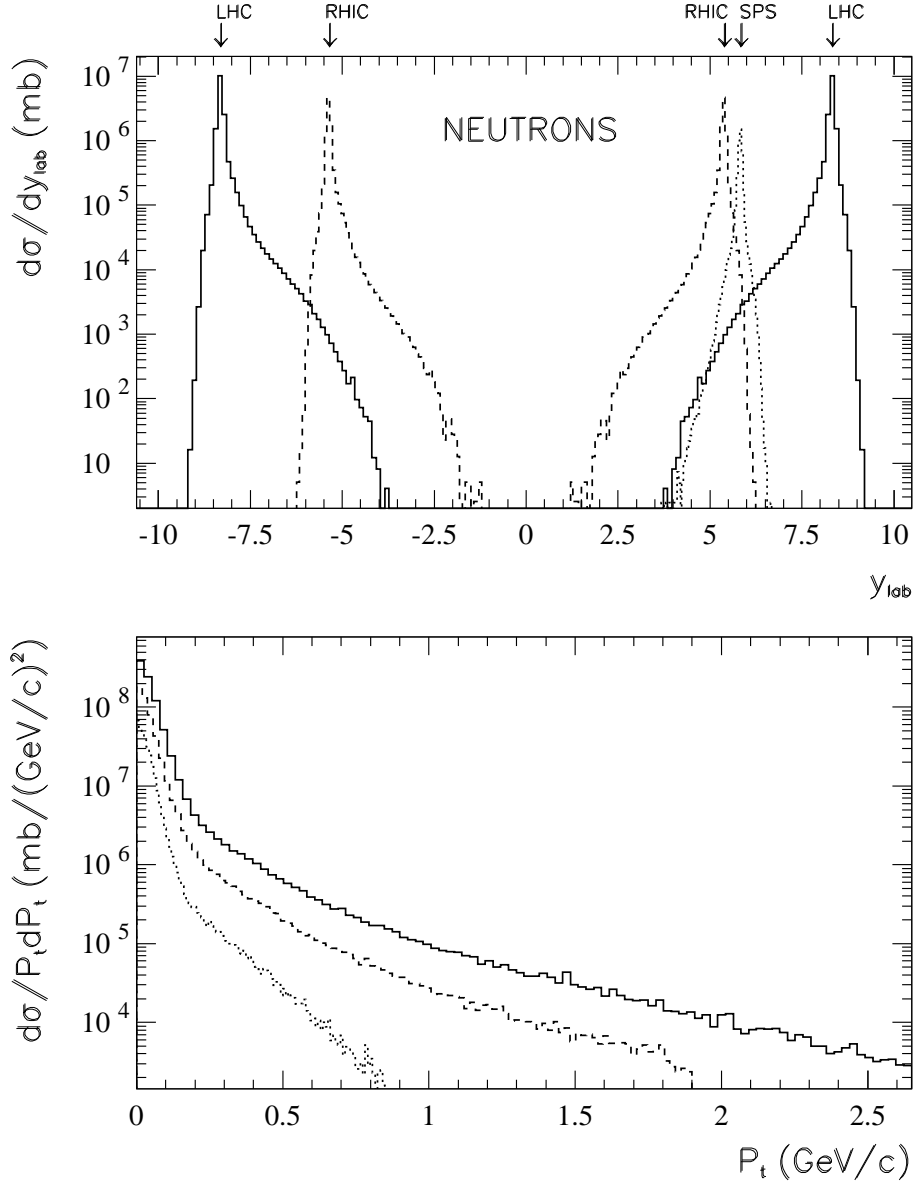}
\caption{ RELDIS predictions for inclusive  rapidity (top panel) and 
transverse momentum (bottom panel)  distributions  
of neutrons emitted in the electromagnetic 
dissociation 
of Pb nuclei at LHC and SPS energies (solid and dotted lines,
respectively) and Au nuclei at RHIC energies (dashed lines).
Rapidity distributions for colliders were reflected respect to the 
value $y_{lab}=0$. Arrows show rapidity of corresponding beams.}
\label{fig:neut}
\end{centering}
\end{figure}

The Lorentz-invariant inclusive differential cross sections,
$d\sigma /dy_{lab}$ and $d\sigma /P_tdP_t$, for neutrons are shown in 
Fig.~\ref{fig:neut}. Obviously, the dominant part of  
neutrons is concentrated near the beam rapidity (see upper panel).
However, the cross section of neutron emission is quite large 
($\sim 0.1-1$ b) even 2-3 units away from the beam rapidity.  
In our model these neutrons are produced due to reactions 
$\gamma N\rightarrow i\pi n$ ($1\leq i\leq 8$) initiated by a high-energy
virtual photon. The recoil neutron in these reactions has 
a chance to keep a great part of the photon energy.

The neutron $P_t$ distributions (lower panel) reveal clearly the presence of
two components associated with fast and slow stages of the reaction. 
The evaporation neutrons are concentrated 
mainly at $P_t\leq 150$ MeV, while the high momentum tail of the 
$P_t$-distribution is formed by early emitted knock-out neutrons.

\begin{figure}[hp]  
\begin{centering}
\epsfxsize=0.95\textwidth
\epsffile{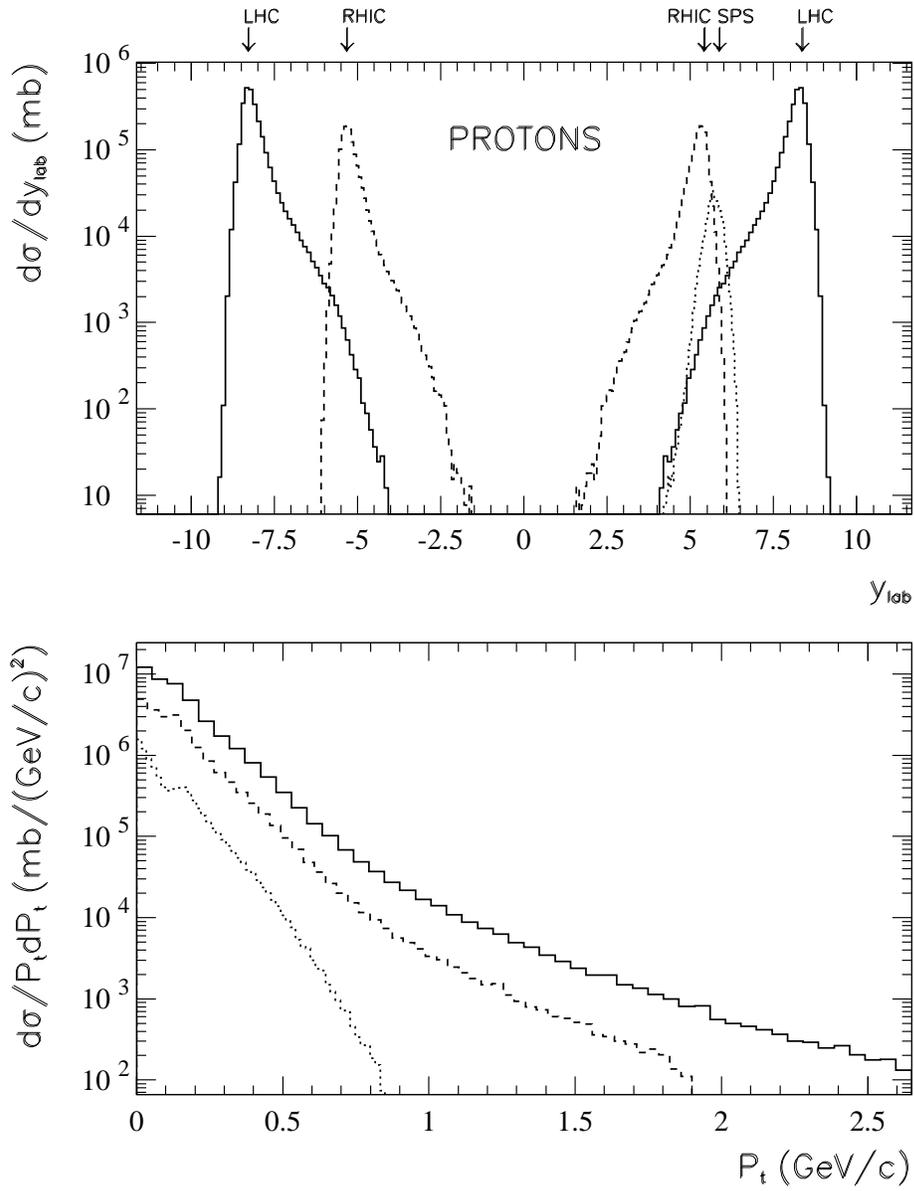}
\caption{ The same as in Fig.~\protect\ref{fig:neut}, but for protons.}
\label{fig:prot}
\end{centering}
\end{figure}

Fig.~\ref{fig:prot} shows the invariant differential cross sections,
$d\sigma /dy_{lab}$ and $d\sigma /P_tdP_t$  for protons. 
Although the proton rapidity distributions are similar to those for neutrons,
their transverse momentum spectra reveal a difference. 
Namely, the evaporation component in the proton spectra is suppressed 
by the Coulomb barrier. Instead of two components, a broad peak is  seen 
at $P_t\sim 150-250$ MeV/c. Its origin can be understood as follows. 
A virtual photon from the region of quasideuteron absorption,
$E_\gamma\sim 30-100$ MeV, produces a pair of nucleons
with approximately equal momenta of 200-300 MeV/c\footnote{Fermi motion is 
neglected in this estimation.}. They are emitted predominantly at 
angles $80^\circ-85^\circ$ with respect to the collision axis in the 
nuclear rest frame 
and therefore the corresponding $P_t$ is approximately $150-250$ MeV/c.
This prediction can be checked by measuring the  $P_t$ distributions of 
protons produced in the ED reactions. Up to now mainly the collective 
excitations of nuclei in the GR region were studied in these reactions 
through the neutron emission~\cite{Aumann1} or fission~\cite{Rubehn}
channels. There were no special attempts to see how virtual photons interact 
directly with individual nucleons.
 
\begin{figure}[hp]  
\begin{centering}
\epsfxsize=0.95\textwidth
\epsffile{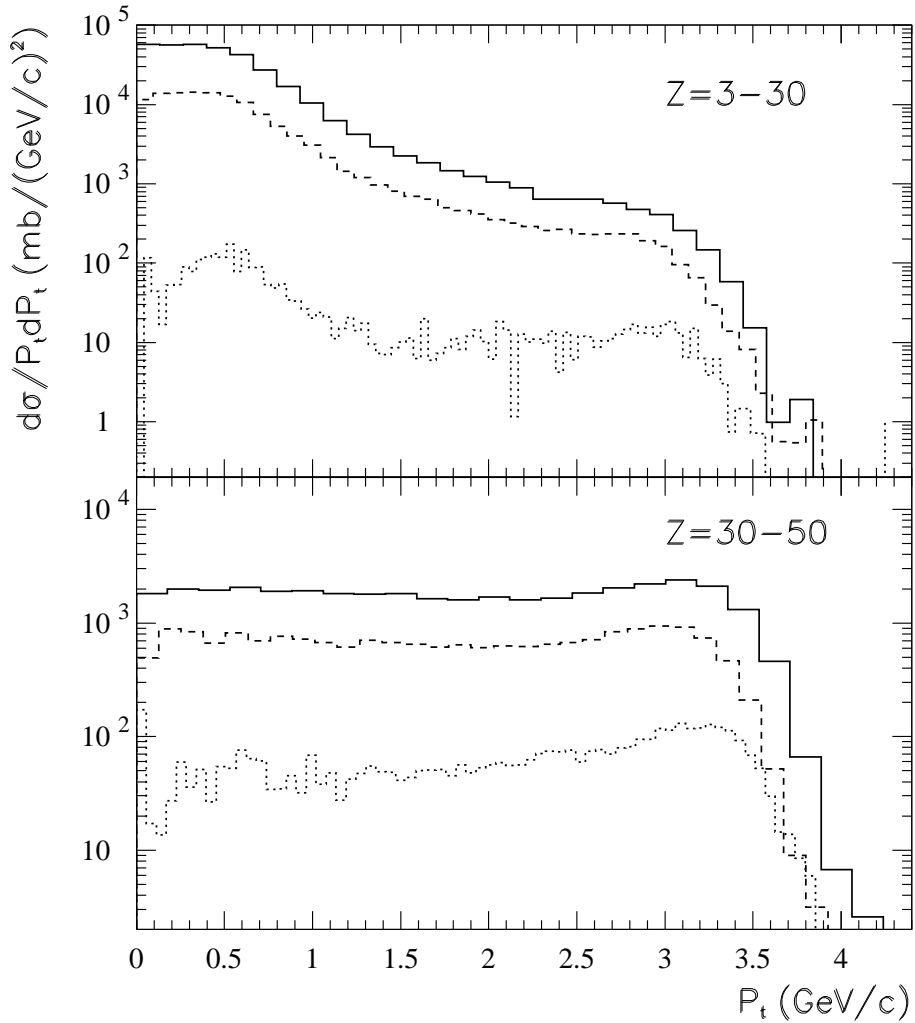}
\caption{ RELDIS predictions for inclusive transverse momentum distributions  
of intermediate mass (top panel) and fission (bottom panel) fragments 
emitted in the electromagnetic 
dissociation 
of Pb nuclei at LHC and SPS energies (solid and dotted lines,
respectively) and Au nuclei at RHIC energies (dashed lines). }
\label{fig:frag}
\end{centering}
\end{figure}

The transverse momentum distributions of 
intermediate-mass ($3\leq Z\leq 30$) and fission  
($30<Z\leq 50$) fragments are shown in Fig.~\ref{fig:frag}.
They are remarkably flat.
All these fragments are produced at late stages of the reaction and 
their characteristic momenta, in the nuclear rest frame, are of about 100 
MeV/c per nucleon. Their rapidity distributions are not shown because they are
extremely narrow and centered around the beam 
rapidities. As shown in Ref.~\cite{Pshenichnov},  
shell structure effects do not play a role at these high excitation 
energies and, 
therefore, fission fragments have a symmetric charge distribution.
The distribution of intermediate-mass fragments is close to a power law.

\subsection{Background for $\gamma\gamma$ processes}

\begin{figure}[hp]  
\begin{centering}
\epsfxsize=0.95\textwidth
\epsffile{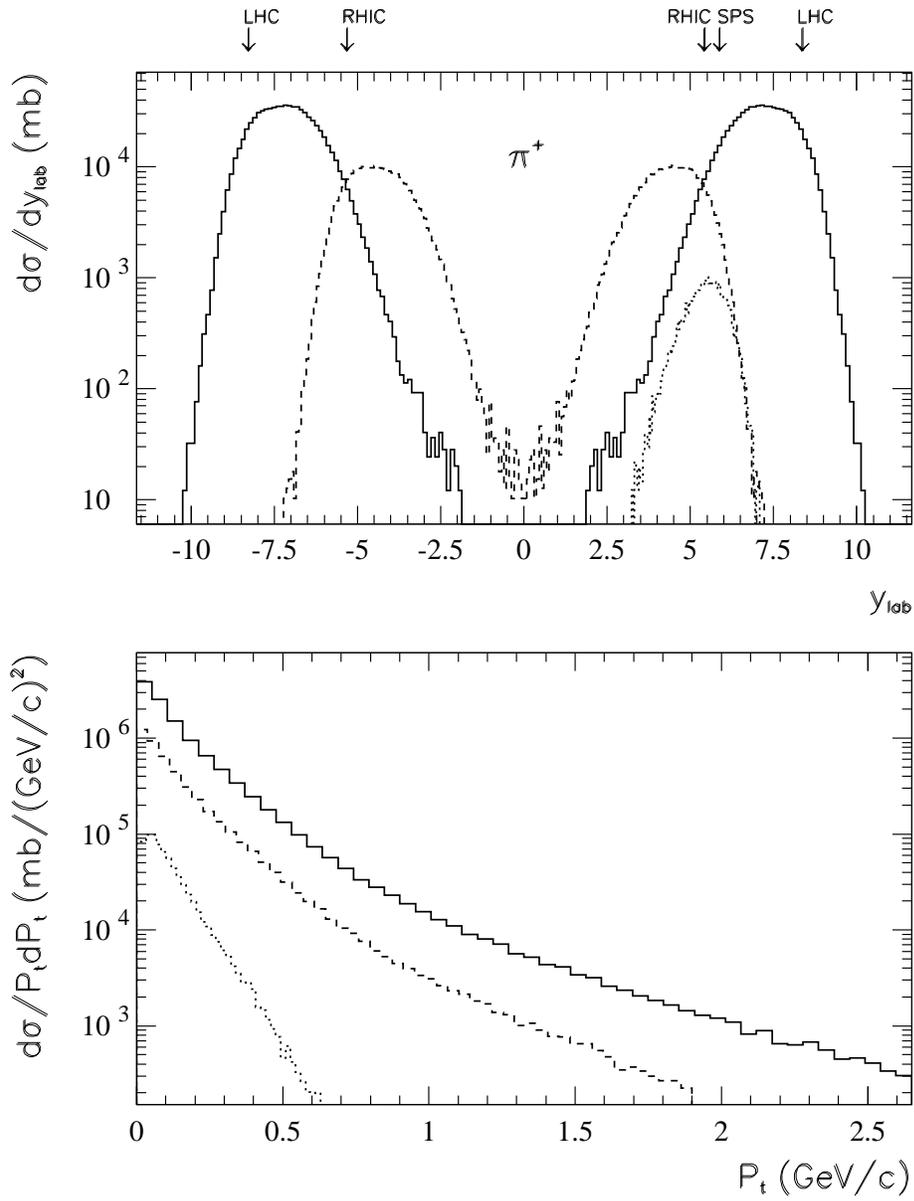}
\caption{ The same as in Fig.~\protect\ref{fig:neut}, but for positive pions.}
\label{fig:pip}
\end{centering}
\end{figure}

\begin{figure}[hp]  
\begin{centering}
\epsfxsize=0.95\textwidth
\epsffile{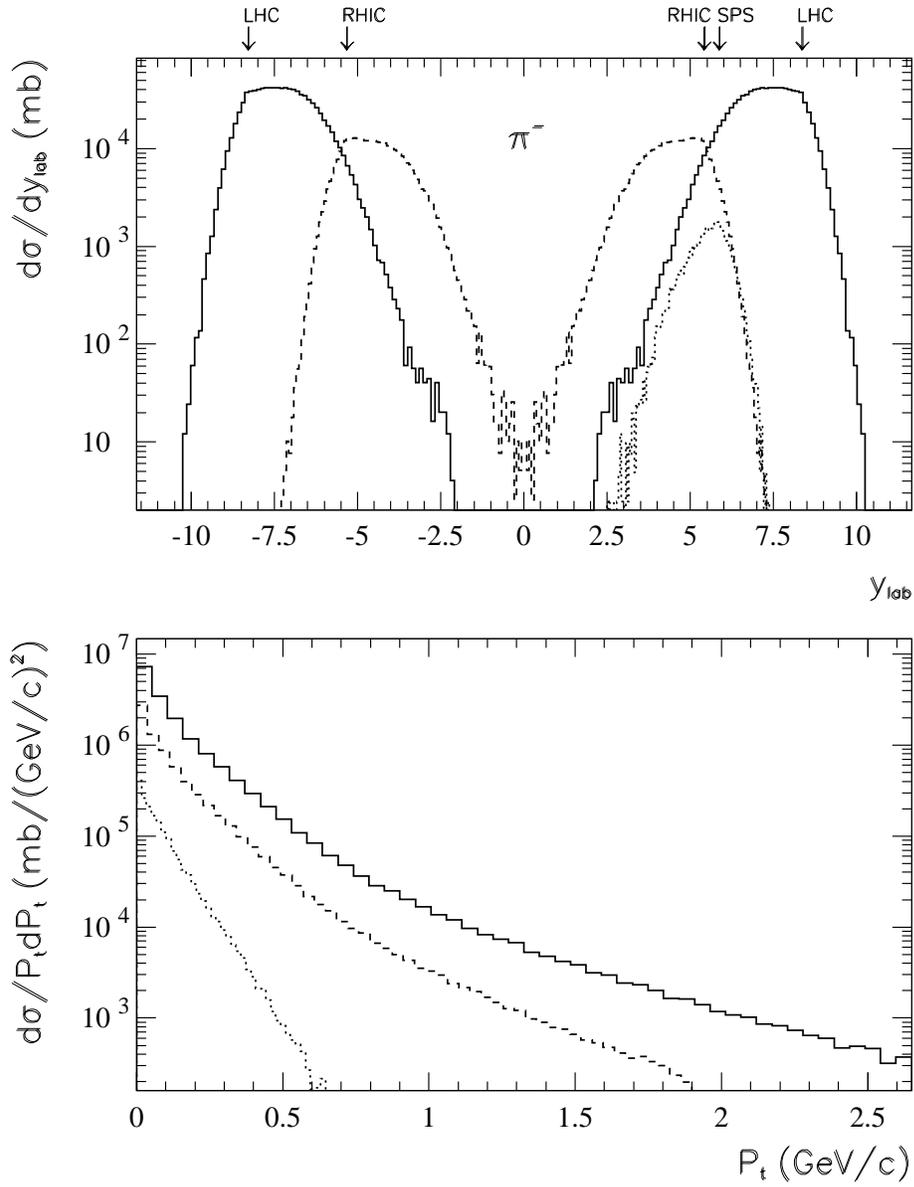}
\caption{ The same as in Fig.~\protect\ref{fig:neut}, but for negative pions.}
\label{fig:pim}
\end{centering}
\end{figure}

In this section we demonstrate that pions produced in the $\gamma A$
process may cause problems in studying $\gamma\gamma$ fusion 
reactions. The predicted distributions of positive and negative pions  from
$\gamma A$ processes are given in
Figs.~\ref{fig:pip} and \ref{fig:pim}, respectively. Both $\pi^+$ and 
$\pi^-$ have very wide rapidity distributions which extend 
toward mid-rapidity up to $|y_{lab}|<2$ for LHC and up to 
$y_{lab}\sim 0$ for RHIC.
One can also notice that some pions have rapidities which exceed the 
beam rapidity ($y_{RHIC}=5.4$, $y_{LHC}=8.3$). This is a peculiarity of the 
multiple pion photoproduction at high energies when a large phase space is
available for pions. Some pions are emitted  backward with respect to 
the virtual photon momentum.

The pions produced  in $\gamma A$  
interactions close to the mid-rapidity region may create complications for
the detection of particles produced from $\gamma\gamma$-fusion 
process~\cite{MGreiner,BaronBaur,Vidovic}.
This can be seen from a simple estimate.
As predicted in Ref.~\cite{Vidovic}, the cross section to produce a
$D^{+}_s$ meson at LHC is 0.144 mb and $d\sigma /dy_{lab}$ is peaked 
at $y_{lab}=0$ so that almost all these mesons will have rapidities within
the interval $|y_{lab}|<2$. The probabilities of $D^{+}_s$  decays
into $\eta\pi^+$, $\omega\pi^+$ or $\rho^0\pi^+$ channels are expected 
at the level of  $1-5\%$,~\cite{PDG}.   Then the cross section to produce 
a $\pi^+$ from the $D^{+}_s$ decay turns out to be below 0.007 mb. On the other
hand, for $\pi^+$ 
from the $\gamma A$ process we predict $d\sigma /dy_{lab}\sim 10$ mb at 
$|y_{lab}|=2$ (see Fig.~\ref{fig:pip}).
These estimations show that it will be very difficult to separate
the products of the $D^{+}_s$ decay from the background of the 
$\gamma A$ process. This does not necessarily mean that the
detection of $D^{+}_s$ is impossible since different kinematical cuts
may be applied to suppress the $\gamma A$ background, as well as other 
decay modes may be used for the  $D^{+}_s$ detection. The considered 
example rather demonstrates the importance of the proper evaluation 
of the $\gamma A$ background in planning
experiments on $\gamma\gamma$ fusion at future colliders.

\section{Conclusions}

In this paper we have further developed the model of electromagnetic
dissociation of nuclei first introduced in Ref.~\cite{Pshenichnov}.
The computer code RELDIS is developed to calculate a wide range of
observable characteristics associated with the ED reactions. This code
includes the dissociation channels with nucleons, pions, fission fragments 
and intermediate mass fragments from nuclear multifragmentation. 
The model calculations are in a good agreement 
with experimental data obtained  for the dissociation of 1 GeV/nucleon 
$^{197}{\rm Au}$ ions. This gives us the basis to put forward predictions 
for the electromagnetic dissociation of ultrarelativistic heavy ions 
at RHIC and LHC energies.
One should bear in mind that the multiplicities of particle produced by 
high-energy virtual photons are rather small as compared with those in
central nuclear collisions. But the reaction rates are large.
Besides the fundamental interest, the ED process may have interesting
applications. As proposed in Ref.~\cite{Baltz},   
the neutron emission in the ED process may be used for luminosity monitoring 
via zero degree calorimeters. Detailed calculations  presented 
in this paper make it possible to eliminate some unnecessary simplifications
in evaluating the neutron transverse momentum 
distributions which were adopted in Ref.~\cite{Baltz}.
Although the advantages of detecting mutual one-neutron dissociation events 
are prominent~\cite{Baltz2}, one should carefully estimate the
contribution of more deep dissociation processes,
as it has been made in the present paper.
The electromagnetic fission of $\rm Au$ and $\rm Pb$ nuclei will be also
noticeable at RHIC and LHC.
An alternative approach to luminosity monitoring would be to measure the 
yields of fission fragments or protons, which can be easily separated from 
the beam nuclei by their different Z/A-ratio.
Using presented above $P_t$ distributions of fission and intermediate mass 
fragments one can estimate the response of forward detectors to such 
fragments. 
The calculated differential distributions of pions should be taken
into consideration in planning experiments on $\gamma\gamma$ fusion at 
CERN~\cite{CMS}.

One can use the predictions of the present model in order to
disentangle different mechanisms of the heavy-ion dissociation, not only 
though the measurement of the Z-dependence of the total cross section,
as it was made in Ref.~\cite{Datz}, but also through the registration of 
some specific exclusive channels (multiple neutron emission, for example). 
Such measurements are needed, in particular, for understanding the origin  
of the correction term in the dissociation cross section found in 
Ref.~\cite{Datz}.

We are grateful to J.J.Gaardh\o{j}e, A.B.Kurepin, M.V.Mebel and L.M.Satarov 
for useful discussions.  
I.N.M., A.S.B. and I.A.P. thank the Niels Bohr Institute 
for the warm hospitality and financial support. 
I.A.P. thanks INTAS for the Young Scientists Fellowship  98-86.
The work was supported partially by the Danish Natural Science Research 
Council and Universities of Russia Basic Research Fund, grant 5347.

\newpage

\begin{centering}
\begin{table}[h]
\caption{Calculated and measured~\protect\cite{Aumann1} cross sections for 
nucleon emission in the reactions
$^{197}{\rm Au}(^{197}{\rm Au},xn)^{197-x}{\rm Au}$ and
$^{197}{\rm Au}(^{197}{\rm Au},pxn)^{196-x}{\rm Pt}$ at 1A GeV.
}
\vspace{0.3cm} 
\begin{tabular}{|c|c|c|c|c|}    
\hline\hline
  &\multicolumn{4}{c|}{\ } \\
  &\multicolumn{4}{c|}{Cross section (mb)} \\ 
  &\multicolumn{4}{c|}{\ } \\
\cline{2-5} 
  & \multicolumn{3}{c|}{\ } &  \\
Channel  & \multicolumn{3}{c|}{Model} & Experiment \\
  & \multicolumn{3}{c|}{\ } &  \\
\cline{2-5}
  & & & & \\  
 & $\sigma^{(1)}_{i}$ & $\sigma^{(2)}_{i}$ 
 & $\sigma_{i}$ & $\sigma_{i}$  \\
  & & & & \\
\hline\hline  
$n$ & 2432 & 74 & 2506 & $3077\pm 200$ \\
$2n$ & 926 & 13 & 939 & $643\pm 105$  \\
$3n$ & 129 & 77 &  206 & $171\pm 26$  \\
\hline 
 Total & & & &\\
$n-3n$ & 3487 & 164 & 3651 & $3891\pm 227$ \\
\hline\hline
$4n$ & 40  & 43   &    83 &   -    \\
$5n$ & 7.7 & 11.6 & 19.3  &   -    \\
$6n$ & 2.2 & 4.4  &   6.6 &   -    \\
\hline 
 Total & & & &\\
$4n-6n$ & 49.9 & 59 & 108.9 & - \\
\hline\hline
$p$ & 1.7 & 0 &  1.7 &   -    \\
$pn$ & 3.2 & 0 & 3.2  &   -    \\
$p2n$ & 2.5 & 0.6 & 3.1 &   -    \\
\hline 
 Total & & & &\\
$p-p2n$ & 7.4 & 0.6 & 8 & - \\
\hline\hline
$p3n$ & 0.8 & 0.7 &  1.5 &   -    \\
$p4n$ & 0.3 & 0.5 &  0.8 &   -    \\
$p5n$ & 0.1 & 0.2 &  0.3   &   -    \\
\hline 
 Total & & & &\\
$p3n-p5n$ & 1.2 & 1.4 & 2.6 & - \\
\hline\hline
\end{tabular}
\label{T1}
\end{table}
\end{centering}

\begin{centering}
\begin{table}[h]
\caption{Total cross sections (barn) of electromagnetic dissociation, 
$\sigma_{ED}$, and contributions from the single, $\sigma^{(1)}$,  and
double, $\sigma^{(2)}$, photon absorption. The values are given for
the present and future heavy-ion beams of Pb and Au nuclei.}
\vspace{0.3cm} 
\begin{tabular}{|c|c|c|c|}    
\hline\hline
  &  &  &  \\
Reaction & $\sigma^{(1)}$ & $\sigma^{(2)}$ & $\sigma_{ED}$  \\ 
  &  &  &  \\ 
\hline\hline  
158A      GeV     & 41.2 & 1.5 & 42.7 \\
$^{208}{\rm Pb}\ on \ ^{208}{\rm Pb}$ &   &  &   \\
\hline
100A+100A GeV     & 93.8 & 2.2  & 96  \\
$^{197}{\rm Au}\ on \ ^{197}{\rm Au}$ &   &  &    \\
\hline
2.75A+2.75A TeV     & 208.7 & 3 & 211.7 \\
$^{208}{\rm Pb}\ on \ ^{208}{\rm Pb}$ &   &  &      \\
\hline\hline
\end{tabular}
\label{T2}
\end{table}
\end{centering}
\begin{centering}
\begin{table}[h]
\caption{Inclusive cross sections (barn) of emission of nucleons, nuclear
fragments and pions in the electromagnetic dissociation of Pb and Au nuclei
calculated for the present and future RHI beams.}
\vspace{0.3cm} 
\begin{tabular}{|c|c|c|c|c|c|c|c|}    
\hline\hline
  &\multicolumn{7}{c|}{\ } \\
Reaction  & 
\multicolumn{7}{c|}{$\sigma^{(1)}_{i}+\sigma^{(2)}_{i}$} \\ 
\cline{2-8} 
\ & $p$ & $n$ & $3\leq Z\leq 30$ & $30<Z\leq 50$ & 
$\pi^+$ & $\pi^-$ & $\pi^0$ \\
\hline\hline  
158A      GeV      & 15.9 & 175  & 0.09 & 0.5   & 1.37  &  2.28 & 2.7  \\
$^{208}{\rm Pb}\ on \ ^{208}{\rm Pb}$    &  &  &  &  &  &  &   \\
\hline
100A+100A GeV      & 104.7 & 686.7& 6.28 & 4.3  & 24 & 30.2 & 33.1  \\
$^{197}{\rm Au}\ on \ ^{197}{\rm Au}$    &  &  &  &  &  &  &   \\
\hline
2.75A+2.75A TeV    & 302.4 & 1853 & 22.3 & 11.9 & 90.4 & 111.6 & 121.4\\
$^{208}{\rm Pb}\ on \ ^{208}{\rm Pb}$    &  &  &  &  &  &  &    \\
\hline\hline
\end{tabular}
\label{T3}
\end{table}
\end{centering}
\begin{centering}
\begin{table}[h]
\caption{Electromagnetic dissociation of Pb and Au nuclei. 
Contributions (\% )
due to the double photon absorption 
to the inclusive cross sections of emission of nucleons, nuclear
fragments and pions.}
\vspace{0.3cm} 
\begin{tabular}{|c|c|c|c|c|c|c|c|}    
\hline\hline
  &\multicolumn{7}{c|}{\ } \\
Reaction  & \multicolumn{7}{c|}
{$\sigma^{(2)}_{i}/(\sigma^{(1)}_{i}+\sigma^{(2)}_{i})$} \\ 
\cline{2-8} 
\ & $p$ & $n$ & $3\leq Z\leq 30$ & $30<Z\leq 50$ & 
$\pi^+$ & $\pi^-$ & $\pi^0$ \\
\hline\hline  
158A      GeV      & 8.1 & 6.7 &14. & 10.5  & 11.6   & 11.1 & 10.1  \\
$^{208}{\rm Pb}\ on \ ^{208}{\rm Pb}$    &  &  &  &  &  &  &   \\
\hline
100A+100A GeV      & 3.7 & 3.3  & 4.2 & 4.  & 5.6 & 5.3  & 5.06  \\
$^{197}{\rm Au}\ on \ ^{197}{\rm Au}$    &  &  &  &  &  &  &   \\
\hline
2.75A+2.75A TeV    & 1.8   & 1.8    & 1.4  & 1.8 & 2.4  & 2.4  & 2.3  \\
$^{208}{\rm Pb}\ on \ ^{208}{\rm Pb}$    &  &  &  &  &  &  &    \\
\hline\hline
\end{tabular}
\label{T4}
\end{table}
\end{centering}

\newpage
%%%%%%%%%%%%%%%%%%%%%%%%%%%%%  Figures %%%%%%%%%%%%%%%%%%%%%%%%%%%%%%%%

%%
%%

%%%%%%%%%%%%%%%%%%%%%%%%%%%%%%%%
\end{document}